\documentclass[fleqn,10pt]{iopart}
\usepackage{graphicx} % Include figure files
\usepackage{amssymb} % for special symbols e.g. \gtrsim, \lesssim
\begin{document}
\title[Non--monotonic orbital velocity profiles]{Non--monotonic orbital velocity profiles around rapidly rotating Kerr--(anti--)de Sitter black holes}

\author{A M\"{u}ller and  B Aschenbach}
\address{Max-Planck-Institut f\"{u}r extraterrestrische Physik, Giessenbachstr. 1, 85748 Garching, Germany}%
\eads{\mailto{amueller@mpe.mpg.de}, \mailto{bra@mpe.mpg.de}}

\begin{abstract}
It has been recently demonstrated that the orbital velocity profile around Kerr black holes in the equatorial 
plane as observed in the locally non--rotating frame exhibits a non--monotonic radial behaviour. We show here 
that this unexpected minimum--maximum feature of the orbital velocity remains if the Kerr vacuum is generalized 
to the Kerr-de Sitter or Kerr--anti--de Sitter metric. This is a new general relativity effect in Kerr spacetimes 
with non--vanishing cosmological constant. Assuming that the profile of the orbital velocity is known, this effect 
constrains the spacetime parameters. 
\end{abstract}
% \submitto{\CQG}
{\vspace{28pt plus 10pt minus 18pt}
     \noindent{\small\rm Received 15 January 2007, in final form 5 April 2007 \par}}
{    \noindent{\small\rm Accepted for publication in \emph{Classical and Quantum Gravity} \par}}
\maketitle

\section{\label{sec:intro}Introduction\protect}
Rotating black holes are described by the Kerr solution of general relativity \cite{Kerr1963}. Matter 
particles can perform stable circular orbits in the equatorial plane around these compact objects as 
long as the orbital radius is greater than the marginally stable orbit \cite{Bardeen1972}. The 
introduction of a locally non--rotating frame (LNRF) offers a fairly easy way for studying particle 
motions in the Kerr geometry \cite{Bardeen1970}. In a sense, the observer's frame co--rotates with 
spacetime thereby cancelling frame--dragging effects as much as possible. Recently, an unexpected 
behaviour of the radial dependence of the orbital LNRF velocity of test masses with a Keplerian angular 
velocity distribution around a rapidly rotating, non--charged black hole has been discovered 
\cite{Aschenbach2004b}: In vacuum spacetime of very fast spinning Kerr black holes with the 
rotational parameter $a>0.9953$, there occurs an unexpected dip of the orbital LNRF velocity as function 
of orbital radius. This minimum--maximum structure of the orbital velocity emerges close to the black 
hole, at radii $r<1.8$. Orbital radii are given in units of the gravitational radius that is defined as 
$r_\mathrm{g}=\mathrm{G}M/\mathrm{c}^2$ with Newton's constant G, vacuum speed of light c and black hole 
mass $M$. We set G=c=$M$=1 throughout the paper for convenience. The gradient of the orbital velocity is 
positive in a small radial range. The radii associated with the local extrema are greater than the innermost 
stable circular orbit (ISCO) and smaller than the last stable orbit at $r=2$ for any value of $a$. Hence, 
this effect occurs always within the ergosphere of a Kerr black hole. The velocity difference between the 
extrema, i.e.\ the slow down amounts approximately to 1\% of c at Thorne's spin limit $a=0.998$ 
\cite{Thorne1974} but is higher for larger black hole spin. The non--monotonic behaviour of the orbital 
LNRF velocity is a pure effect of general relativity and has been overlooked up to 2004. A follow--up 
investigation has shown that the non--monotonic behaviour remains even for non--Keplerian distributions 
of angular momentum ($l={\rm const}$) in the Kerr vacuum \cite{Stuchlik2005}. The critical value of the black 
hole spin that guarantees the emergence of the effect is higher in this case, i.e.\ $a>0.99979$.

\section{\label{sec:effect}Kerr spacetimes with non--vanishing $\Lambda$\protect}
% added in revised version, 1st referee report (1) - begin
The motivation for this work is the question whether the non--monotonicity occurs also in more general Kerr
spacetimes or might actually been removed. 
% added in revised version, 1st referee report (1) - end
In the present work, the LNRF orbital velocity is generalized for Kerr--de Sitter and Kerr--anti--de Sitter 
spacetimes, i.e.\ including the cosmological constant $\Lambda$. As for the Kerr spacetime proper,  
we use the Boyer--Lindquist form \cite{Boyer1967} to study the motion of a particle. Then, the line 
element of the ordinary Kerr geometry holds
\begin{equation} \label{eq:Kerr}
ds^{2}=-\alpha^{2}dt^{2}+\tilde\omega^{2}(d\phi-\omega dt)^{2}+\rho^{2}/\Delta \ dr^{2}+\rho^{2} d\theta^{2},
\end{equation}
with the functions (${\rm G}={\rm c}=M=1$)
\begin{eqnarray}
\alpha & = & \rho\sqrt{\Delta}/\Sigma, \\
\Delta & = & r^2-2r+a^2, \\
\rho^2 & = & r^2+a^2\,\cos^2\theta, \\
\Sigma^2 & = & (r^2+a^2)^2-a^2\Delta\sin^2\theta, \\
\omega & = & 2ar/\Sigma^2, \\
\tilde{\omega} & = & \Sigma\sin\theta/\rho,
\end{eqnarray}
where $M$ and $a$ denote black hole mass and spin, respectively. 

The line element for the Kerr--de Sitter (KdS, $\Lambda>0$) or Kerr--anti--de Sitter (KadS, $\Lambda<0$) 
metric, respectively, is significantly more complicated \cite{Carter1968,Gibbons1977,Stuchlik1991}
% Carter 1968 added in revised version 
% 3: Gibbons & Hawking, 4: Stuchlik & Calvani 
% \citep{Gibbons1977, Stuchlik1991}
\begin{eqnarray}
ds^{2} & = & -\frac{\Delta_r}{\chi^2\,\rho^2}\,\left(dt-a\,\sin^2\theta\,d\phi\right)^2 +\frac{\Delta_\theta\,\sin^2\theta}{\chi^2\,\rho^2} \nonumber\\
& & \times\left[a\,dt-(r^2+a^2)\,d\phi\right]^2 +\rho^2\,\left(\frac{dr^2}{\Delta_r}+\frac{d\theta^{2}}{\Delta_\theta}\right),
\end{eqnarray}
where we have the generalized functions
\begin{eqnarray}
\Delta_r & = & (r^2+a^2)\,(1-\frac{1}{3}\,\Lambda r^2)-2r, \\
\Delta_\theta & = & 1 + \frac{1}{3}\,\Lambda a^2\cos^2\theta, \\
\chi & = & 1 + \frac{1}{3}\,\Lambda a^2.
\end{eqnarray}
$\Lambda$ denotes the cosmological constant. These functions reduce to the ordinary Kerr geometry 
by setting $\Lambda =0$:
\begin{eqnarray}
\Delta_r & = & r^2+a^2-2r\equiv\Delta, \\
\Delta_\theta & = & 1, \\
\chi & = & 1.
\end{eqnarray}
Now, we aim to cast the KdS/KadS line element into a form analogous to equation~(\ref{eq:Kerr})
\begin{equation}
ds^{2}=-\alpha_{\Lambda}^{2}\,dt^{2}+\tilde\omega_{\Lambda}^{2}\,(d\phi-\omega_{\Lambda}\,dt)^{2}+\rho^2\,\left(\frac{dr^2}{\Delta_r}+\frac{d\theta^{2}}{\Delta_\theta}\right),
\end{equation}
where $\alpha_{\Lambda},\,\tilde\omega_{\Lambda},\,\omega_{\Lambda}$ are the generalizations 
of the functions $\alpha,\,\tilde\omega,\,\omega$, respectively. Indeed, this is possible and we find
\begin{eqnarray}
\omega_{\Lambda} & = & \frac{a\,[(r^2+a^2)\Delta_\theta-\Delta_r]}{\Delta_\theta\,(r^2+a^2)^2-\Delta_r\,a^2\sin^2\theta}, \\
\alpha^2_{\Lambda} & = & \frac{\Delta_r\,\Delta_\theta\,\rho^2}{\chi^2\,\left[\Delta_\theta\,(r^2+a^2)^2-\Delta_r\, a^2\sin^2\theta\right]}, \\
\tilde{\omega}^2_{\Lambda} & = & \frac{\sin^2\theta}{\chi^2\,\rho^2}\left[\Delta_\theta\,(r^2+a^2)^2-\Delta_r\,a^2\sin^2\theta\right].
\end{eqnarray}
Note that the first expression represents the generalization of the frame--dragging frequency, 
$\omega_{\Lambda}$. Considering that the shift vector in the Kerr geometry satisfies $\beta^\phi=-\omega$ 
in Boyer--Lindquist form \cite{Thorne1986}, it is interesting to investigate its generalization, 
$\beta^\phi_\Lambda=-\omega_\Lambda$. It is easy to show that $\omega_\Lambda$ is larger for increasing
values of the cosmological constant, i.e.\ the frame--dragging effect becomes stronger with increasing 
$\Lambda$.

Further, it is even possible to reverse frame--dragging with a sufficiently low and negative $\Lambda$. 
In this case, the radial position where the reversion takes place is determined by the root of 
$\omega_\Lambda$. 

With these generalizations it is straightforward to compute the orbital velocity component relative 
to the LNRF for Kerr spacetimes with non--zero $\Lambda$ just analogous to the ordinary Kerr geometry, 
see e.g. \cite{Mueller2004}:
% 6: Bardeen (1970)
% 7: BPT1972
\begin{eqnarray} \label{eq:velKdS}
\fl v^{(\phi)}_{\Lambda} & = & \tilde{\omega}_{\Lambda}\ \frac{\Omega-\omega_{\Lambda}}{\alpha_{\Lambda}} \nonumber\\
\fl & = & \frac{\sin\theta\,[\Delta_\theta\,(r^2+a^2)^2-\Delta_r\,a^2\sin^2\theta]}{\rho^2\,\sqrt{\Delta_r\,\Delta_\theta}}\,\left\{\Omega-\frac{a\,[(r^2+a^2)\Delta_\theta-\Delta_r]}{\Delta_\theta\,(r^2+a^2)^2-\Delta_r\,a^2\sin^2\theta}\right\}.
\end{eqnarray}
We investigate this completely general radial profile of the orbital LNRF velocity by specifying the 
angular velocity $\Omega$ for two cases: Keplerian or constant specific angular momentum rotation.
Both distributions are motivated by astrophysical situations. Keplerian angular momentum distributions can 
be found in accretion flows described by the standard disc model \cite{Shakura1973}. Torus solutions exhibit constant angular 
momentum distributions \cite{Abramowicz1978}.

\subsection{\label{sec:kep}Keplerian distribution of the specific angular momentum\protect}
The non--monotonic orbital velocity profile has originally been discovered for prograde orbits. A similar effect has not been
found by us for retrograde orbits. Hence we assume a prograde ($+$) 
Keplerian angular velocity distribution of the orbiting test particles that 
satisfies
\begin{equation} \label{eq:OmegKep}
\Omega=\Omega_{\rm K}^+=\frac{1}{\sqrt{r^3}+a},
\end{equation}
we can compute the velocity profile from equation~(\ref{eq:velKdS}). The result is shown in figure~1 which
demonstrates that the minimum--maximum structure depends on $\Lambda$. As visible in the plot, a cosmological 
constant close to zero self--consistently approaches the ordinary case for the Kerr geometry. Further, the strength of 
the slow--down effect increases as $\Lambda$ decreases, i.e.\ Kerr--anti--de Sitter ($\Lambda < 0$) exhibits a stronger 
effect than Kerr--de Sitter ($\Lambda > 0$).
% change Fig. 01 in revised version, 1st referee report (4) + (5)
\begin{figure}
\begin{center}
\includegraphics[width=8cm]{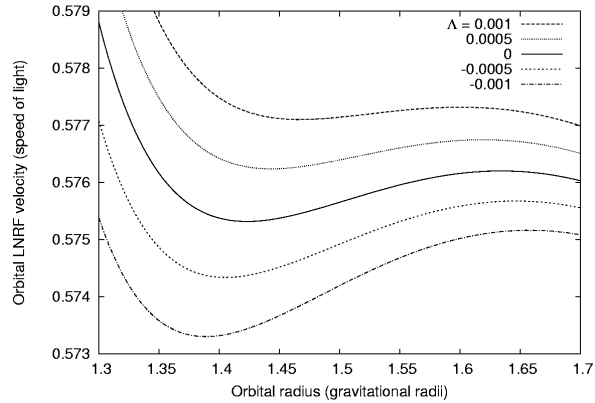}
\caption{\label{fig:01} Orbital LNRF velocity profiles around a black hole with $a=0.996$ for Keplerian orbits 
in the equatorial plane ($\theta=\pi/2$) and different values of the cosmological constant, 
$\Lambda=10^{-3},\,5\times 10^{-4},\,0,\,-5\times 10^{-4},\,-10^{-3}$.
The orbital velocity is given in units of c ($\mathrm{G}=\mathrm{c}=\mathrm{M}=1$).}
\end{center}
\end{figure}

\subsection{\label{sec:lconst} Constant specific angular momentum distribution\protect}
Particles around black holes may not follow Keplerian orbits but show some distribution 
of the specific angular momentum $l$, e.g.\ $l={\rm const}$. In this case, the general expression for the 
angular velocity in the Kerr geometry can be specified for the Kerr--(anti--)de Sitter geometry
\begin{equation} \label{eq:genOmeg}
\Omega=-\frac{l g_{\rm tt}+g_{{\rm t}\phi}}{l g_{{\rm t}\phi}+g_{\phi\phi}} =\omega_\Lambda+\frac{\alpha_\Lambda^2}{\tilde{\omega}_\Lambda^2}\frac{l}{1-\omega_\Lambda l}.
\end{equation}
The specific angular momentum $l$ has to be chosen in a range between marginally stable and 
marginally bound orbits, i.e.\ $l_{\rm ms}\leq l\leq l_{\rm mb}$, see \cite{Abramowicz1978}. 
Fixing $l$ in this interval again reveals distinct minimum--maximum 
structures in the velocity profile. Figure~2 displays the velocity profile computed from 
equations~(\ref{eq:velKdS}) and (\ref{eq:genOmeg}) for a Kerr black hole with $a=0.9999$ and 
$l=l(r=1.0785)=l_{\rm ms}$. 
In figure~2 the values for $a$ and $l$ are chosen as an example and to make our work comparable 
to the work of \cite{Stuchlik2005}. For $\Lambda=0$ their result is confirmed (solid curve). 
The sign of $\Lambda$ can be easily determined by comparing the profile's position relative to 
the case $\Lambda=0$. Similar to the Keplerian distribution, the trend remains: a negative 
$\Lambda$ amplifies the slow--down effect.
\begin{figure}
\begin{center}
\includegraphics[width=8cm]{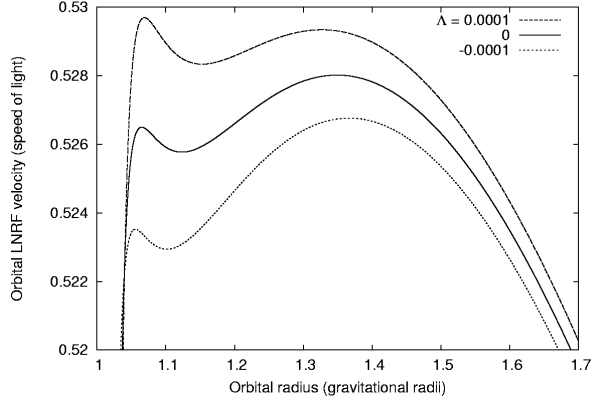}
\caption{\label{fig:02} Orbital LNRF velocity profiles around a black hole with $a=0.9999$ for 
$l={\rm const}$ orbits in the equatorial plane ($\theta=\pi/2$) and different values of  the cosmological 
constant, $\Lambda=10^{-4},\,0,\,-10^{-4}$.}
\end{center}
\end{figure}

\section{\label{sec:int}An interplay of black hole spin and $\Lambda$\protect}
The examples discussed so far demonstrate an interesting interplay of black hole spin $a$ and 
$\Lambda$. We have investigated this more deeply since the interaction seems to control the 
occurrence and modulation depth of the minimum--maximum structure. We consider only the 
equatorial plane ($\theta=\pi/2$) and compute the gradient of the 
orbital LNRF velocity for prograde Keplerian angular momentum distribution, $\Omega=\Omega^+_{\rm K}$, 
and find
\begin{eqnarray}
\fl \frac{\partial v^{(\phi)}_\Lambda}{\partial r} = [\,a^6\Lambda r (9-\Lambda r^2)+2a^5\Lambda r^{3/2}(\Lambda r^3-3)+a^4(18+27r-6\Lambda r^2+9\Lambda r^3-4\Lambda^2 r^5) \nonumber\\
+2a^3\sqrt{r}\,(-9-27r+12\Lambda r^3+3\Lambda r^4 +\Lambda^2 r^6) \nonumber\\
-3a^2 r^2(30-12r+6\Lambda r^2+3\Lambda r^3+\Lambda^2 r^5)+6a r^{5/2}(27-15r+3\Lambda r^3+\Lambda r^4) \nonumber\\
+9r^5(1-\Lambda r^2)]\slash \{2\,\sqrt{3r}\,(a+r^{3/2})^2\,\sqrt{a^2\,(3-\Lambda r^2)-r(6-3r+\Lambda r^3)} \nonumber\\
\times[\,a^2\,(\Lambda r^2-3)+r(6-3r+\Lambda r^3)]\}.
\end{eqnarray}
Analogously, the gradient of the orbital LNRF velocity can be analytically computed by plugging 
equation~(\ref{eq:genOmeg}) into equation~(\ref{eq:velKdS}), i.e. for constant angular momentum 
distributions. This gradient satisfies
\begin{eqnarray}
\fl \frac{\partial v^{(\phi)}_\Lambda}{\partial r} = \{\,l[-a^6\Lambda^2r^3+a^5\Lambda^2r^3l-a^4(-18+\Lambda^2 r^5+3\Lambda r^2\,(5+2r)) \nonumber\\
+a^3 l\,(-18+\Lambda^2 r^5+3\Lambda r^2\,(5+r))-3a^2 r\,(18-9r+3r^2 +3\Lambda r^3+2\Lambda r^4) \nonumber\\
+3alr\,(18-12r+3\Lambda r^3+\Lambda r^4)-9\,r^4(r-3)]\}\slash \nonumber\\
\{\sqrt{a^2-2r+r^2-a^2\Lambda r^2/3-\Lambda r^4/3} \nonumber\\
\times[\,a^4\Lambda r -a^3\Lambda l r+3 r^3-a l (6+\Lambda\,r^3)+a^2(6+3r+\Lambda r^3)]^2\}.
\end{eqnarray}
We study both gradient equations in the following and start with the prograde Keplerian angular velocity. The roots of 
the gradient determine the radial positions of the local extrema. In general, the existence and the positions 
of extrema are controlled by $a$ and $\Lambda$. With decreasing $\Lambda$ the slow--down effect is enhanced 
for fixed and sufficiently high black hole spin. A gradient with only one root has a point of inflexion 
in the velocity profile, whereas more roots yield minima and maxima of the slow--down structure. For each given 
$\Lambda$, there exists a critical value of the black hole spin $a_\mathrm{c}$ that is associated with a gradient 
vanishing at one specific radius. For spin values greater than $a_\mathrm{c}$, there exists a radial range where the 
gradient becomes positive. In the Kerr vacuum ($\Lambda=0$), it is $a_\mathrm{c}>0.9953$ for Keplerian 
orbiters. However, generalizing to non--zero $\Lambda$ it is even possible that a high value of $\Lambda$ is 
associated with a lower value of critical spin. A graphical analysis of the gradient for fixed $\Lambda$ 
delivers the critical spins $a_\mathrm{c}$ beyond which the minimum--maximum structure appears. The critical 
spin $a_\mathrm{c}$ scales linearly with $\Lambda$ for $|a_\mathrm{c}|<1$ (figure~3). The study for ordinary 
Kerr black holes by Aschenbach \cite{Aschenbach2004b} is confirmed and generalized to cases with non--zero cosmological 
constant.

Unexpectedly, a similar linear correlation between $a_\mathrm{c}$ and $\Lambda$ can be found for $l={\rm const}$ angular
momentum distributions (figure~4). Both the slope and the offset of $a$ for $\Lambda=0$ differ slightly. The critical 
spin grows more steeply and the offset is smaller for Keplerian angular momentum distributions. Further, the linear 
correlation is constrained to a narrow interval of the cosmological constant. We find in our analysis that the 
minimum--maximum structure appears only for 
$-0.001\lesssim\Lambda\lesssim 0.00017$. 
% Fig. 03 in revised version is old Fig. 04 of subm version
\begin{figure}
\begin{center}
\includegraphics[width=8cm]{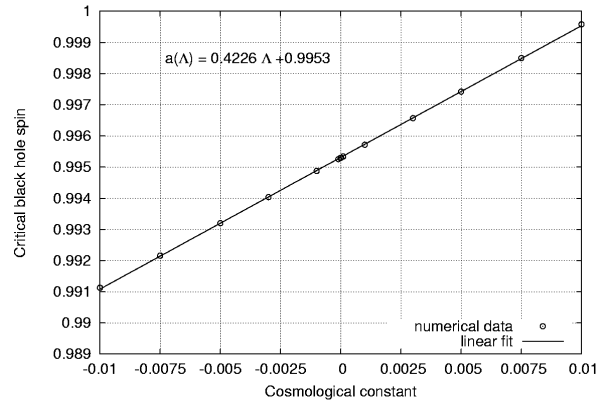}
\caption{\label{fig:03} Critical spin values as function of $\Lambda$ for Keplerian angular momentum distribution. The 
limiting values for the black hole spin $a_\mathrm{c}$ are plotted as function of $\Lambda$ for Keplerian distributions 
of the specific angular momentum. There appears to be an almost positive linear correlation between $a_\mathrm{c}$ 
and $\Lambda$. For spins larger than $a_\mathrm{c}$ the slow--down effect occurs. However, the linear 
relation breaks down for $\Lambda>0.011$, because of $a>1$.}
\end{center}
\end{figure}
\begin{figure}
\begin{center}
\includegraphics[width=8cm]{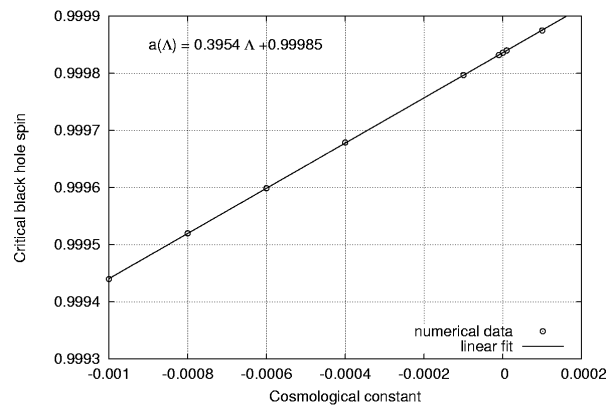}
\caption{\label{fig:04} Critical spin values as function of $\Lambda$ for $l={\rm const}$ angular momentum distribution. The 
limiting values for the black hole spin $a_\mathrm{c}$ are plotted as function of $\Lambda$ for $l={\rm const}$ specific angular momentum 
distributions. Similar to the Keplerian case an positive linear correlation between $a_\mathrm{c}$ and $\Lambda$ remains. The 
analysis of the orbital velocity profiles shows that the the minimum--maximum structure occurs only for
$-0.001\lesssim\Lambda\lesssim 0.00017$.}
\end{center}
\end{figure}

\section{\label{sec:cosmo}Discussion\protect}
The slow--down effect occurs for rapidly spinning black holes, $a\gtrsim 0.99$. Both observations and theory suggest the existence 
of rapidly spinning black holes. Rapid rotation of supermassive black holes is suggested by the observation of broad iron K X--ray 
lines \cite{Tanaka1995}, and by flares observed from the 
galactic centre black hole \cite{Genzel2003,Porquet2003,Aschenbach2004a}. Black hole theory suggests that supermassive 
black holes are endowed with high angular momentum due to the black hole growth history \cite{Shapiro2005,Volonteri2005}. 
Jet launching mechanisms such as the Blandford--Znajek process \cite{Blandford1977} also involve rapidly spinning black 
holes. General relativistic magnetohydrodynamics simulations support this idea because strong outflows are driven by 
Poynting fluxes near black holes only for high spins \cite{Krolik2005,McKinney2006}.

If the minimum--maximum structure around a fast spinning black hole could be detected, its radial profile constrains 
both black hole spin and $\Lambda$. 

If black hole mass and spin were known from observations, e.g.\ from quasi--periodic oscillations, it might be possible 
to constrain the cosmological constant. But only extraordinarily high values of $\Lambda$ would produce a significant 
change in the feature as demonstrated in the numerical examples. The cosmic concordance model \cite{Spergel2006} suggests 
that $\Lambda$ is by many orders of magnitudes too small to enforce a significant change in the minimum--maximum structure. 
Currently, there is no hope to detect this for the Kerr black hole candidates. On the other hand, we would like to stress 
that theory of dark energy physics is limited by a number of uncertainties allowing a plethora of models. The extremes 
include proposals advocating for a cosmological constant that was significantly higher in earlier cosmological 
epochs, e.g.\ involving false vacua \cite{Volovik2005}, or even that there is no need for a cosmological constant 
because density inhomogeneities drive a modification of cosmic expansion \cite{Kolb2005}. Another model involves a 
modification of 4D gravity \cite{Dvali2000}. The new GR effect outlined here might be exploited to probe general 
relativity for the strong field case. It is currently hard to say whether this new effect has any practical implication 
for cosmological studies.

\section{\label{sec:conc}Conclusions\protect}
We conclude that a very fast spinning black hole ($a>0.991$ for $\Lambda<0$ or $a>0.9953$ for $\Lambda>0$) causes 
a slow down of the orbital velocity at distances within two gravitational radii. Closer to the rotating black hole 
the orbital velocity steeply increases again -- just 
as expected from Newtonian physics. It has been demonstrated in this paper that the 'GR slow--down effect' survives 
as the Kerr black hole is immersed into a $\Lambda$ fluid which is described by the Kerr--de Sitter or Kerr--anti--de 
Sitter solution. This is true for Keplerian distributions of angular momenta as well as for distributions with constant 
angular momenta. The parameter study reveals a trend that the slow--down effect is more pronounced for negative values 
of $\Lambda$ in both cases, the Keplerian and $l={\rm const}$ orbiters. The minimum--maximum structure occurs for black hole 
spins close to 1 and fairly high $\Lambda$. This is a new GR effect in Kerr--(anti--)de Sitter spacetimes. 

\ack
We thank an anonymous referee who helped us to improve the paper.

\end{document}